\documentclass[12pt]{iopart}

\usepackage{epsf,iopams,graphicx}

\begin{document}

\title[Fidelity amplitude of the scattering matrix]
{Fidelity amplitude of the scattering matrix in microwave
cavities}

\author{R Sch\"afer$^1$, T Gorin$^2$,
        T H Seligman$^3$, and H-J St\"ockmann$^1$}

\address{$^1$ Fachbereich Physik, Philipps-Universit\"at Marburg, Renthof 5,
D-35032 Marburg, Germany}

\address{$^2$  Max-Planck-Institut f\" ur Physik komplexer Systeme,
               N\" othnitzer Str. 38, D-01187 Dresden, Germany}

\address{$^3$ Centro de Ciencias F\'{\i}sicas, Universidad Nacional
Aut\'{o}noma de M\'{e}xico, Campus Morelos, C.~P. 62251,
Cuernavaca, Morelos, M\'{e}xico}



\begin{abstract}
The concept of fidelity decay is discussed from the point of view of
the scattering matrix, and the {\it scattering fidelity}
is introduced as the
parametric cross-correlation of a given S-matrix element, taken in
the time domain, normalized by the corresponding autocorrelation
function. We show that for chaotic systems, this quantity
represents the usual fidelity amplitude, if appropriate ensemble and/or energy
averages are taken. We present a microwave experiment where the
scattering fidelity is measured for an ensemble of chaotic systems.
The results are in excellent agreement with random matrix theory for
the standard fidelity amplitude. The only parameter, namely the
perturbation strength could be determined independently from level
dynamics of the system, thus providing a parameter free agreement
between theory and experiment.
\end{abstract}



\ead{rudi.schaefer@physik.uni-marburg.de}

\maketitle

\newcommand {\xb} {{\bf x}}
\newcommand {\xbd} {{\bf x^\dag}}
\newcommand {\yb} {{\bf y}}
\newcommand {\ybd} {{\bf y^\dag}}
\newcommand {\zn} {{\bf z}_n}
\newcommand {\znd} {{\bf z}_n^{\bf\dag}}
\newcommand {\ree} {R_\phi\left(E_1,E_2\right)}

\newcommand {\bet}{B}
\newcommand {\gam}{C}

\newcommand{\eff}{{\rm eff}}
\newcommand{\inter}{{\rm int}}
\newcommand{\tot}{{\rm tot}}

\newcommand{\la}{\langle}
\newcommand{\ra}{\rangle}
\newcommand{\lla}{\left\langle}
\newcommand{\rra}{\right\rangle}

\newcommand{\TG}{{\scriptsize\bf\quad [TG]}}

\section{Introduction}

The stability of quantum systems under perturbation of the
dynamics has attracted a considerable amount of attention in
recent years. From a practical point of view this was motivated by
a surge of activity in quantum information, where this question is
crucial; from a more abstract point of view the implications
for ``quantum chaos'' have been studied~\cite{Peres84}. Fidelity,
as a measure for this stability, has become a standard benchmark 
for the reliability of quantum information processing
\cite{nielsen}. Following reference~\cite{pro03b}, we shall
consider two unitary operators $U^{\prime}(t)$ and
$U(t)$ for the perturbed and unperturbed time evolution,
respectively. The fidelity amplitude for some initial state
$\psi(0)$ is defined as
\begin{equation}\label{eq:fid}
f(t) = \left< \psi(0) \right| U^\dagger(t) \, U^{\prime}(t)
   \left| \psi(0)\right>
\end{equation}
and the fidelity by $F(t)=|f(t)|^2$. In other words, the fidelity
amplitude is the cross-correlation function of an initial state
$\psi(0)$ evolving under two different dynamics, namely $U(t)$ and
$U'(t)$. It can be reinterpreted as the autocorrelation function
of the same initial state evolving under the echo operator $M(t)=
U^\dagger(t) \, U^{\prime}(t)$ which propagates the system forward
and backward in time.

The reason why the fidelity or fidelity amplitude are so attractive
is that it gives a measure of the average sensitivity to
perturbations. In quantum optics there are proposals for experiments
to measure directly the fidelity amplitude. The idea is to study the
centre of mass motion of a two level atom in a potential being
slightly different for the first and the second level. Preparing the
atom in a superposition of the two levels and measuring the decay of
the off-diagonal matrix element of the $2\times2$ density matrix,
one obtains the fidelity amplitude of the center of mass motion of
the atom in the external potential~\cite{GPSS04,GCZ97, bienert}.

The motivation for
the present work comes from recent fidelity studies using
microwaves~\cite{SSGS04} and elastic waves \cite{Weaver}. In both
types of experiments the waves are coupled to the systems by means
of antennas or transducers, and reflection or transmission
measurements are performed. The antennas correspond to external
channels, i.\,e. scattering theory is needed to interpret the
experiments. It thus becomes evident that a generalized concept of
fidelity is called for based on the elements of the scattering
matrix.

To study the sensitivity of the S-matrix to perturbations we
therefore propose to define the scattering fidelity of an S-matrix
element as the cross-correlation function of a matrix element of a
perturbed and an unperturbed S-matrix taken in the time-domain and
normalized by the corresponding autocorrelation functions. We
shall see that this adequately describes the experimental
situation, by presenting an experiment with two flat microwave
cavities, both chaotic but one with and one without bouncing-ball
states.

We show that scattering fidelity is equivalent to the fidelity
amplitude~(\ref{eq:fid}) for chaotic dynamics and weak coupling; in
that case furthermore all scattering fidelities are equivalent and
we therefore can improve averages by averaging over channels. The
possibility to replace or improve ensemble averages for S-matrix
elements was theoretically discussed in reference~\cite{nemes} and
applied in reference~\cite{jung} to evaluate the distribution of
absolute values of diagonal and off-diagonal S-matrix elements in a
problem with many open channels.  We then compare the results of our
experiment with the result for a random matrix model developed in
\cite{gor04}. Indeed the experimental results are reproduced without
a fit parameter if the somewhat tedious task is carried out to
determine the strength of the perturbation independently from level
dynamics.

In the next section we define scattering fidelity and its weak
coupling limit, then we describe the experimental setup (section 3). In
section 4 we present a semi-classical evaluation of the perturbation
strength for our setup, followed by a presentation of the
experimental results and a comparison with random matrix theory
(section 5). Particular attention is payed to the effect of bouncing
ball states. In the conclusions (section 6), we discuss situations
where we expect the scattering fidelity to differ from the fidelity
amplitude, and we consider possible experiments.

\section{\label{M} Scattering fidelity}

Here, we analyse the stability of a scattering system under
perturbation of the Hamiltonian. This leads us to the definition
of the scattering fidelity for individual matrix elements of the
scattering matrix. Once the new quantity is defined, we ask, under
which circumstances that quantity approximates the standard
fidelity, equation~(\ref{eq:fid}).

Consider a scattering system which can be perturbed in a controlled
way,  e.\,g. by changing an external parameter. Suppose $S'(E)$ and
$S(E)$ are the scattering matrices corresponding to the perturbed
and unperturbed system, respectively. We denote a cross-correlation
function of two scattering matrix elements by
\begin{equation}
C[S^*_{ab},S'_{cd}](E) = \lla S^*_{ab}(E')\, S'_{cd}(E'+E)\rra -
   \lla S^*_{ab}(E')\rra\; \lla S'_{cd}(E'+E)\rra \; ,
\label{M:Cfun}\end{equation}
where the brackets denote an energy window and/or
an ensemble average. Even though the scattering matrix is defined
in terms of stationary solutions to a dynamical system, we may use
its Fourier transform $\hat S(t)$, to obtain a time dependent
quantity. Due to the convolution theorem~\cite{FTbook} such
correlation functions are natural candidates for the construction
of an open systems' analogue to the fidelity amplitude.
\begin{equation}
\hat C[S^*_{ab},S'_{ab}](t) = \int\rmd E\; \rme^{2\pi\rmi\, Et}\;
   C[S^*_{ab},S'_{ab}](E) \propto \la \hat S_{ab}(t)^*\; \hat S'_{ab}(t)\ra \; .
\label{M:FCfun}\end{equation} This quantity is dominated by the
decay of the autocorrelations, and we therefore include a
normalization into the definition of the scattering fidelity
amplitude:
\begin{equation}
f_{ab}(t)= \frac{\la \hat S_{ab}(t)^*\; \hat S'_{ab}(t) \ra}
   {\sqrt{\la |\hat S_{ab}(t)|^2\ra\; \la |\hat S'_{ab}(t)|^2\ra}} \; .
\label{eq:def_fscat}\end{equation}
Note that a spectral average taken in equation~(\ref{M:Cfun}), is
equivalent to an appropriate smoothing of the time signals
(see reference~\cite{Weaver}).
%


In what follows, we assume that the scattering system can be
described by the effective Hamiltonian approach~\cite{Mah69}.
Then, for chaotic systems it holds that the
definition~(\ref{eq:def_fscat}) approaches the usual fidelity
amplitude in the weak-coupling limit. To show that, we consider
the following parametrization of the scattering matrix
\begin{equation}
\fl S_{ab} (E) = \delta_{ab}
   - \rmi\;{V^{(a)}}^\dagger\; \frac{1}{E-H_{\rm eff}}\; V^{(b)} \qquad
H_{\rm eff}= H_\inter - (\rmi/2)\, \left[ V\, V^\dagger +
\Gamma_W\right] \; , \label{M:Smat}\end{equation} where $V^{(a)}$
is the column vector of $V$ corresponding to the scattering
channel $a$. For later use, we introduce the absorption width
$\Gamma_W$, which is simply a scalar in the present context. This
is equivalent to model absorption with infinitely many
perturbatively coupled channels, whose partial widths add up to
$\Gamma_W$~\cite{Sch03a}.

Consider two slightly different Hamiltonians $H_\inter=
H_\inter(\lambda_1)$ and $H'_\inter= H_\inter(\lambda_2)$, and the
corresponding S-matrices as defined in~(\ref{M:Smat}) and denoted
by $S_{ab}$ and $S'_{ab}$. The Fourier transform of $S'_{ab}$
reads:
\begin{equation}
\fl \hat{S'}\!_{ab}(t) = -\rmi\int\rmd E\; \rme^{-2\pi\rmi\, E\, t}\;
      {V^{(a)}}^\dagger\; \frac{1}{E- H'_{\rm eff}}\; V^{(b)}
 = -2\pi\; \theta(t)\; {V^{(a)}}^\dagger\; \rme^{-2\pi\rmi\,
      H'_{\rm eff}\, t}\; V^{(b)} \; ,
\end{equation}
and similarly for $S_{ab}$. 
Here, $\theta(t)$ is the Heaviside function. The
effective Hamiltonians $H'_{\rm eff}$ and $H_{\rm eff}$ differ only
in their Hermitian part which is $H'_\inter$ or $H_\inter$,
respectively. Without restricting generality (otherwise, we must
perform an Engelbrecht-Weidenm\" uller transformation~\cite{EngWei73}),
we may assume that $V= \sum_c
\sqrt{w_c}\, \la v_c|$, where $|v_1\ra,\ldots|v_M\ra$ are
ortho-normal vectors in the Hilbert space of the closed system.
For the product of two Fourier transformed S-matrix elements, as needed in
equation~(\ref{M:FCfun}), we get:
\begin{equation}
\fl \hat S_{ab}(t)^*\, \hat S'_{ab}(t) = 4\pi^2\theta(t)\, w_a
w_b\;
   \rme^{-2\pi\Gamma_W t}\; \la v_b|
    \rme^{2\pi\rmi\, H_\inter\, t -\pi VV^\dagger\, t} \; | v_a\ra \;
    \la v_a|\; \rme^{-2\pi\rmi\, H'_\inter\, t -\pi VV^\dagger\, t}\;
    | v_b\ra \; .
\label{eq:tauprime}\end{equation} The most unpleasant difference
to the usual definition of a fidelity amplitude is the projector
$|v_a\ra \, \la v_a|$ between forward and backward propagation.
However, under chaotic dynamics, and if $|v_a\ra$ and $|v_b\ra$
are statistically independent, the energy/ensemble average
converts that projector into the unit matrix (multiplied by
$1/L$). We then obtain:
\begin{equation} \label{eq:avg_va}
\lla \hat S_{ab}(t)^*\, \hat S'_{ab}(t)\rra = \frac{4\pi^2}{L}\;
\theta(t)\; w_a\, w_b\;
   \rme^{-2\pi\Gamma_W t}\; \la v_b|\, U_{\rm eff}^\dagger(t)\,
   U'_{\rm eff}(t)\, |v_b\ra \; ,
\end{equation}
where $U_{\rm eff}(t)$ and $U'_{\rm eff}(t)$ are the sub-unitary
propagators for the respective internal Hamiltonians with channel
coupling.
This looks formally like a fidelity amplitude for the effective
Hamiltonians of the system. However, even without any perturbation,
this quantity is not constant, but
yields the
autocorrelation function (i.\,e. the power spectrum). This is the
motivation to define the {\em scattering fidelity} $f_{ab}$ for both
diagonal and off-diagonal S-matrix elements via the relation
\begin{equation} \label{eq:defScatFid}
\lla \hat S_{ab}(t)^*\, \hat S'_{ab}(t)\rra = f_{ab}(t) \; \lla |\hat
S_{ab}(t)|^2 \rra \; ,
\end{equation}
or, more symmetrically, by equation~(\ref{eq:def_fscat}).

In the limit of weak coupling to the antennas, the scattering
fidelity $f_{ab}(t)$  approaches the fidelity of the closed system.
More rigorously, it must be required that the time scale for decay
associated to the coupling $V\, V^\dagger$ is small as compared to
the time scale for the decay of the scattering fidelity. In this
case the wall absorption, described by $\Gamma_W$, is the only
remaining effect leading to:
\begin{equation} \label{eq:avg_va_wcl}
\lla \hat S_{ab}(t)^*\, \hat S'_{ab}(t)\rra \sim
 \frac{4\pi^2}{L}\; \theta(t) \;  w_a\, w_b \, \rme^{-2\pi\Gamma_W t} \;
  f(t) \sim  f(t)\; \lla |\hat S_{ab}(t)|^2\rra \; ,
\end{equation}
where $f(t)$ is the fidelity
amplitude of the closed system with respect to the initial state
$|v_b\ra$. This argumentation assumes that $|v_a\ra$ and $|v_b\ra$
are statistically independent, in particular $a\neq b$. The result
holds, however, for the case $a=b$ as well, provided that
$|v_a\ra$ is effectively a random state in the eigenbasis of
$H_\inter$.

For chaotic dynamics, we have shown that in the limit of small
antenna coupling, the scattering fidelity $f_{ab}(t)$ approaches the
fidelity of the closed system. In this case, there is no restriction
on the strength of the perturbation. However, an alternative derivation
may be based on the rescaled
Breit-Wigner approximation~\cite{Gor02a,Sch03a}. That line of
argument is presented in \ref{app}. It is valid in the perturbative
regime, only, but permits larger antenna couplings (as long as the
rescaled Breit-Wigner approximation remains valid). The experimental
results presented below have been obtained in either one of these
regimes.
%

\section{Experimental setup}

Since a detailed description of the general experimental technique
can be found e.\,g. in \cite{Kuh00b}, we concentrate on the aspects
relevant in the present context. Reflection and transmission
measurements have been performed in a flat microwave cavity, with
top and bottom plate parallel to each other. The cavity is
quasi-two-dimensional for frequencies $\nu < \nu_{max}=c/(2h)$,
where $h$ is the height of the billiard. In this regime there is a
complete equivalence between the stationary wave equation and the
corresponding stationary Schr\"odinger equation, where the $z$
component of the electric field corresponds to the quantum
mechanical wave function,
\begin{equation}
(\Delta+ E)\; \Psi(x,y) = 0 \; , \quad E = \left(\frac{2\pi\,
\nu}{c}\right)^2 \; , \label{eq:Enu}\end{equation} with Dirichlet
boundary conditions.

To describe the experiments adequately, it is essential to take
into account that the microwave system is in fact an open system.
The microwave field in the cavity is constantly fed via one
antenna, while the second antenna and, most importantly, the
non-ideal cavity walls act as sinks. Scattering theory,
originally developed in nuclear physics \cite{Mah69}, can be
applied directly to open microwave billiards as
well~\cite{Stoe99}, if absorption is properly taken into
account~\cite{Sch03a}. With $H_\inter$ being the Hamiltonian of
the closed billiard, and two antennas projecting into the
resonator at the positions $\vec r_1$ and $\vec r_2$, the
scattering matrix is precisely of the form given in equation~(\ref{M:Smat}).
In the present context,
$V^{(a)}$ corresponds to the antenna
at position $\vec r_a$.
For antenna diameters small compared to the wavelength the matrix elements
$V_{ja}$ are proportional to $\psi_j(\vec r_a)$, the wave
functions of the unperturbed system at the antenna positions.
Absorption is included by adding the width $\Gamma_W$ (as a
scalar) to the imaginary part of $H_{\rm eff}$.

Both antennas consisted of copper wires with a diameter of
$1\,$mm, projecting $3\,$mm into the resonator; their positions in
the billiard were fixed for all measurements. An Agilent 8720ES
vector network analyzer was used to determine the complete
S-matrix. Measurements were taken in the frequency range from $3$
to $17\,$GHz with a resolution of $0.5\,$MHz.

\begin{figure}
 \centerline{ \includegraphics[width=0.48\textwidth]{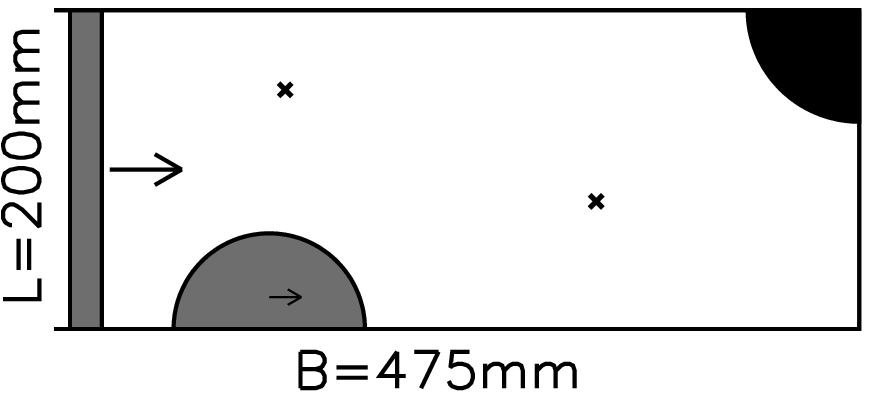}
  \includegraphics[width=0.48\textwidth]{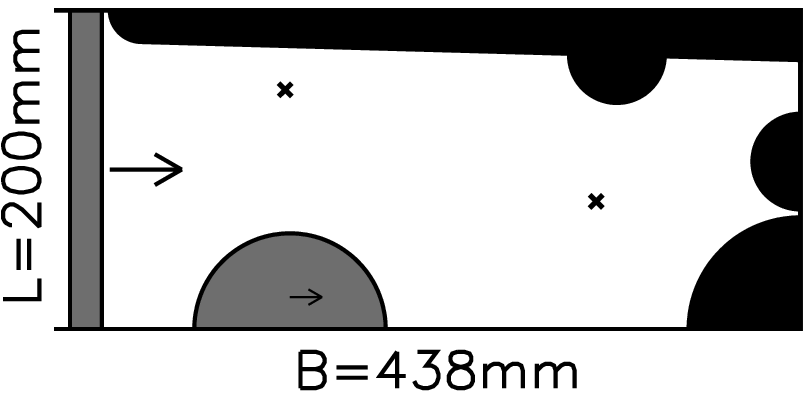}  }
\caption{Geometry of the billiards. In the right billiard bouncing
balls have been avoided by inserting additional elements (see text
for dimensions).} \label{fig:setup}
\end{figure}

The geometries of the billiards studied in this paper are shown in
figure~\ref{fig:setup}. One billiard consists of a rectangular
cavity of length $L=475$\,mm, width $B=200$\,mm and height
$h=8$\,mm, a quarter-circle insert of radius $R_1=70$\,mm, and a
half-circle insert of radius $R_2=60$\,mm placed at the lower
boundary. The position of the latter was changed in steps of
20\,mm to get  15 different systems for the ensemble average. The
perturbation of the system was achieved by moving the left wall in
steps of 0.2\,mm. For each of the 15 positions of the half-circle
we performed 11 measurements for different perturbation strengths.

The other billiard consists of the same rectangular cavity, but
with length $L=438$\,mm. It also shares the quarter-circle insert
of radius $R_1=70$\,mm, and on the lower boundary the half-circle
insert of radius $R_2=60$\,mm, which again was moved to realize an
ensemble of 15 systems. Additional elements were inserted into the
billiard to avoid bouncing-ball resonances: two half-circle
inserts with radius $R_3=30$\,mm, and a wedge
on the upper boundary. Again the perturbation of the
system was realized by moving the left wall in 10 steps of
0.2\,mm.

The classical dynamics for the geometry of the billiards is
dominantly chaotic, and since these are time-reversal invariant
systems, we are going to compare the experimental results with
random matrix predictions for the Gaussian orthogonal ensemble.

\section{Perturbation parameter for shifting of a billiard wall}

In billiard systems the parameter variation is not due to a change
of the Hamiltonian, but of the boundary condition. It was shown in
chapter 5 of reference \cite{Stoe99} that both situations are
equivalent for the case that the parameter variation in the
billiard is due to a shift of a straight wall. In this case the
matrix element of the equivalent perturbation is given by
\begin{equation}
(H_1)_{nm}=l \int_0^L \frac{\partial \psi_n(0,y)}{\partial x}
\frac{\partial \psi_m(0,y)}{\partial x} dy \, ,
\end{equation}
where $l$ is the shift of the wall (in $x$-direction) and $L$ is
the length of the shifted wall. It follows for the perturbation
strength:
\begin{eqnarray}
\fl
\lambda^2 & = & \left< \left[ (H_1)_{nm} \right]^2 \right> \nonumber \\
\fl & = & l^2 \int_0^L dy_1 \int_0^L dy_2 \left< \frac{\partial
\psi_n(0,y_1)}{\partial x} \frac{\partial \psi_n(0,y_2)}{\partial
x} \right> \left< \frac{\partial \psi_m(0,y_1)}{\partial x}
\frac{\partial \psi_m(0,y_2)}{\partial x} \right>
\label{eq:lambda_int}
\end{eqnarray}
This is true for $n\neq m$, but again the result can be easily
generalized to the case $n=m$. The averages can be calculated by
means of Berry's conjecture of the superposition of random plane
waves~\cite{Berry77}. Close to a straight wall with Dirichlet
boundary conditions, this yields
\begin{eqnarray}
\fl
\left< \psi_n(x_1,y_1) \psi_n(x_2,y_2) \right> \nonumber \\
\fl \quad = \frac{1}{A} \left[ J_0 \left( k
\sqrt{(x_1-x_2)^2+(y_1-y_2)^2} \right) - J_0 \left( k
\sqrt{(x_1+x_2)^2+(y_1-y_2)^2} \right) \right]
\end{eqnarray}
for the spacial correlation function, where $A$ is the billiard
area.
Inserting this into equation (\ref{eq:lambda_int}) we obtain for
the case $n\neq m$:
\begin{eqnarray*}
\lambda^2 & = & \left< \left[ (H_1)_{nm} \right]^2 \right> \\[2ex]
& = & \frac{4k^2l^2}{A^2} \int_0^L dy_1 \int_0^L dy_2
\frac{1}{(y_1-y_2)^2}  \left[ J_0^{\prime} \left( k |y_1-y_2| \right) \right]^2  \\[2ex]
& = & \frac{4k^2l^2}{A^2} \int_0^L dy_1 \int_{-y_1}^{L-y_1} dy_2
\frac{1}{y_2^2}  \left[ J_0^{\prime} \left( k y_2 \right) \right]^2   \\[2ex]
& \approx & \frac{4k^2l^2}{A^2} \int_0^L dy_1
\int_{-\infty}^{\infty} dy
\frac{1}{y^2}  \left[ J_0^{\prime} ( y ) \right]^2    \\[2ex]
& = & \frac{4k^2l^2L}{A^2} \frac{8}{3\pi}
\end{eqnarray*}
The approximation works well for large wave numbers $k$.
Since the mean level distance is normalized to one, we can insert
$A=4\pi$ for the area of the billiard, and finally end up with:
\begin{equation} \label{eq:scaling}
\lambda^2 = \frac{2 L}{3 \pi^3} k^3 l^2
\end{equation}

The variance of the diagonal matrix elements, $\left< \left[
(H_1)_{nn} \right]^2 \right>$, yields a value twice as large.
Exactly the same expression was obtained by Leb\oe uf and Sieber
in a completely different approach~\cite{Leb99} using periodic
orbit theory and the ergodicity assumption. It is interesting to
note that two seemingly unrelated assumptions, one on the Gaussian
distribution of wave function amplitudes, and the other on the
ergodicity of long periodic orbits, yield identical results.

\section{Experimental results}

\subsection{Correlation function and fidelity amplitude}

\begin{figure}
\centerline{
\includegraphics[width=0.48\textwidth]{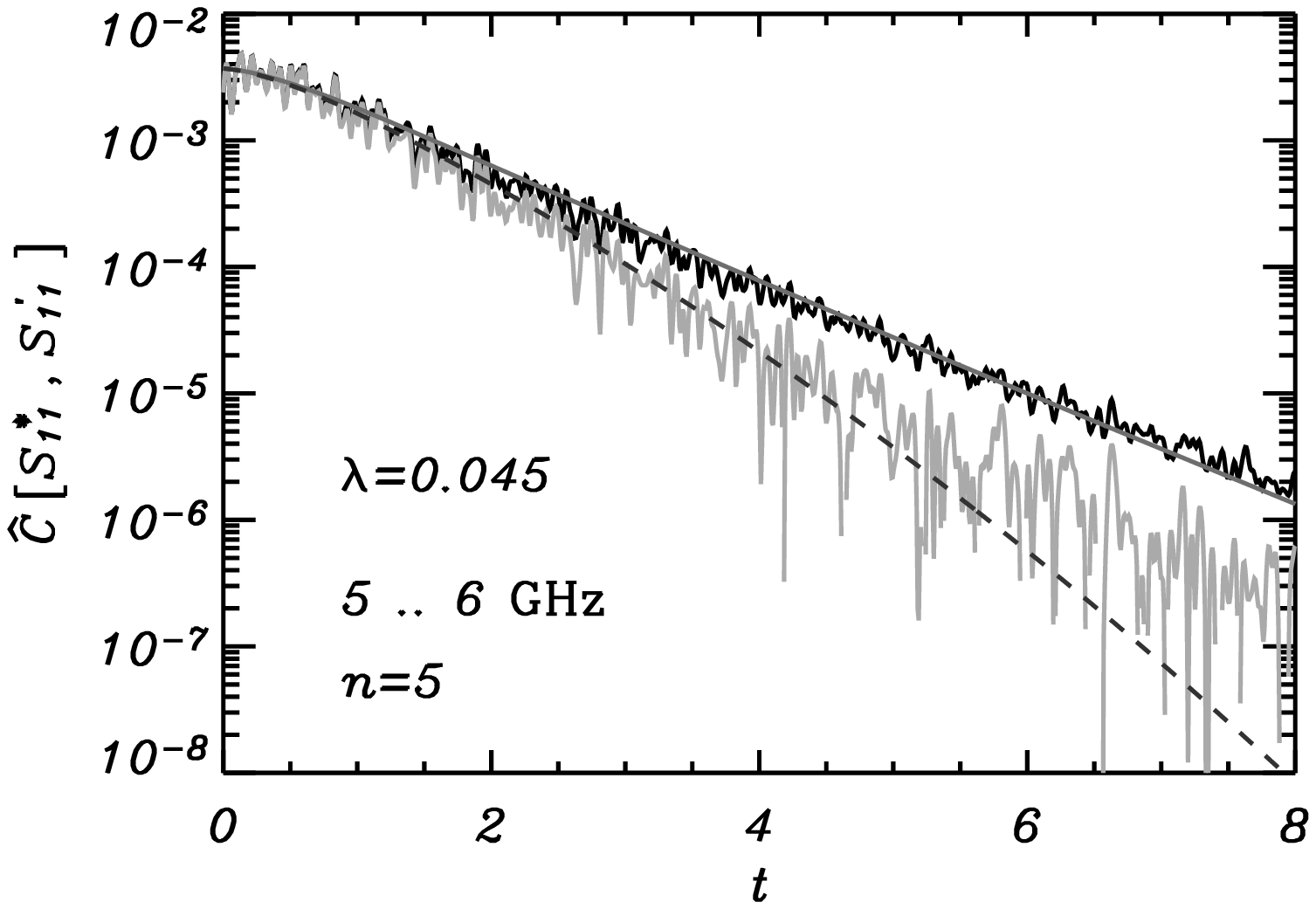}
 \includegraphics[width=0.48\textwidth]{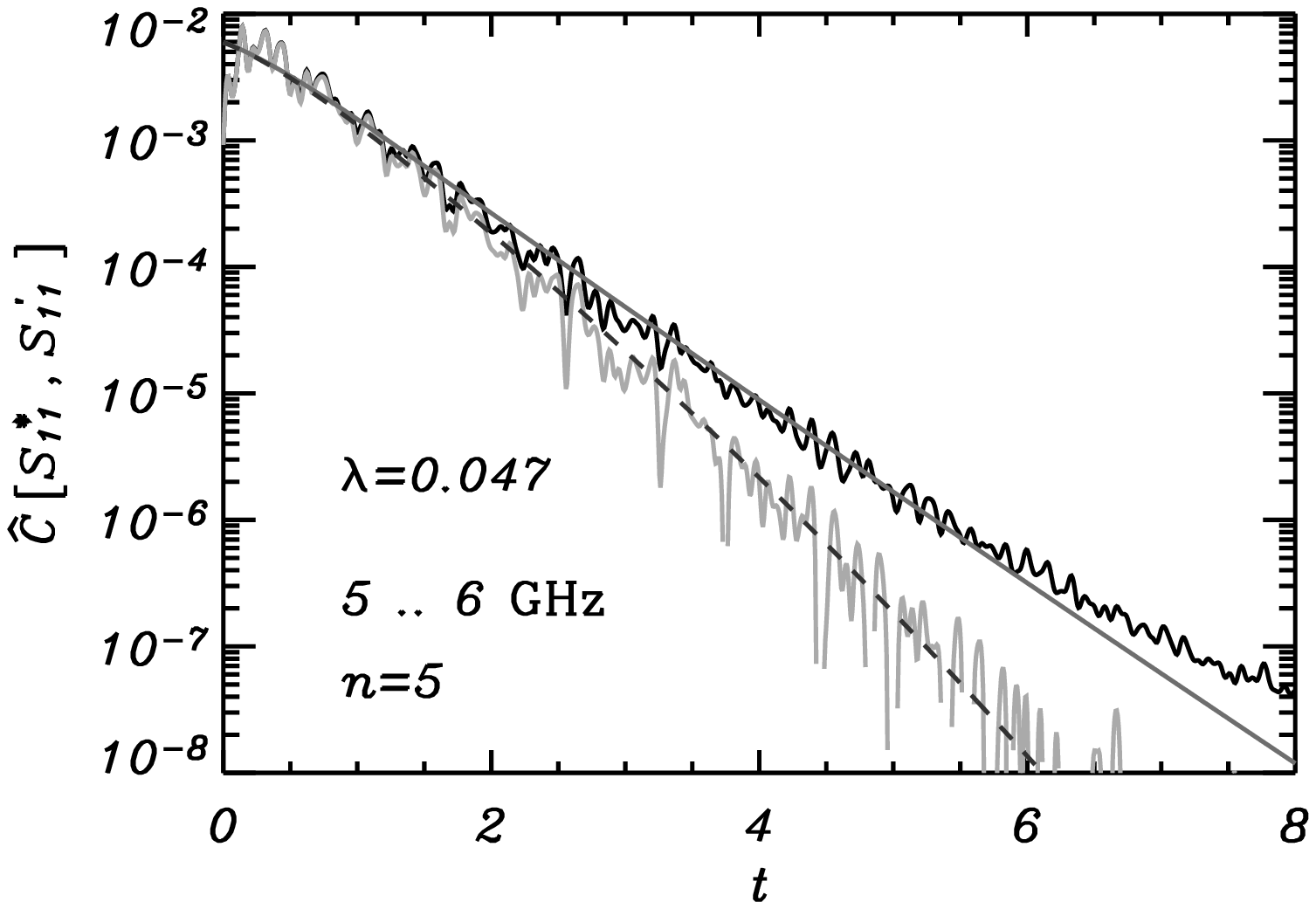} }

\centerline{
\includegraphics[width=0.48\textwidth]{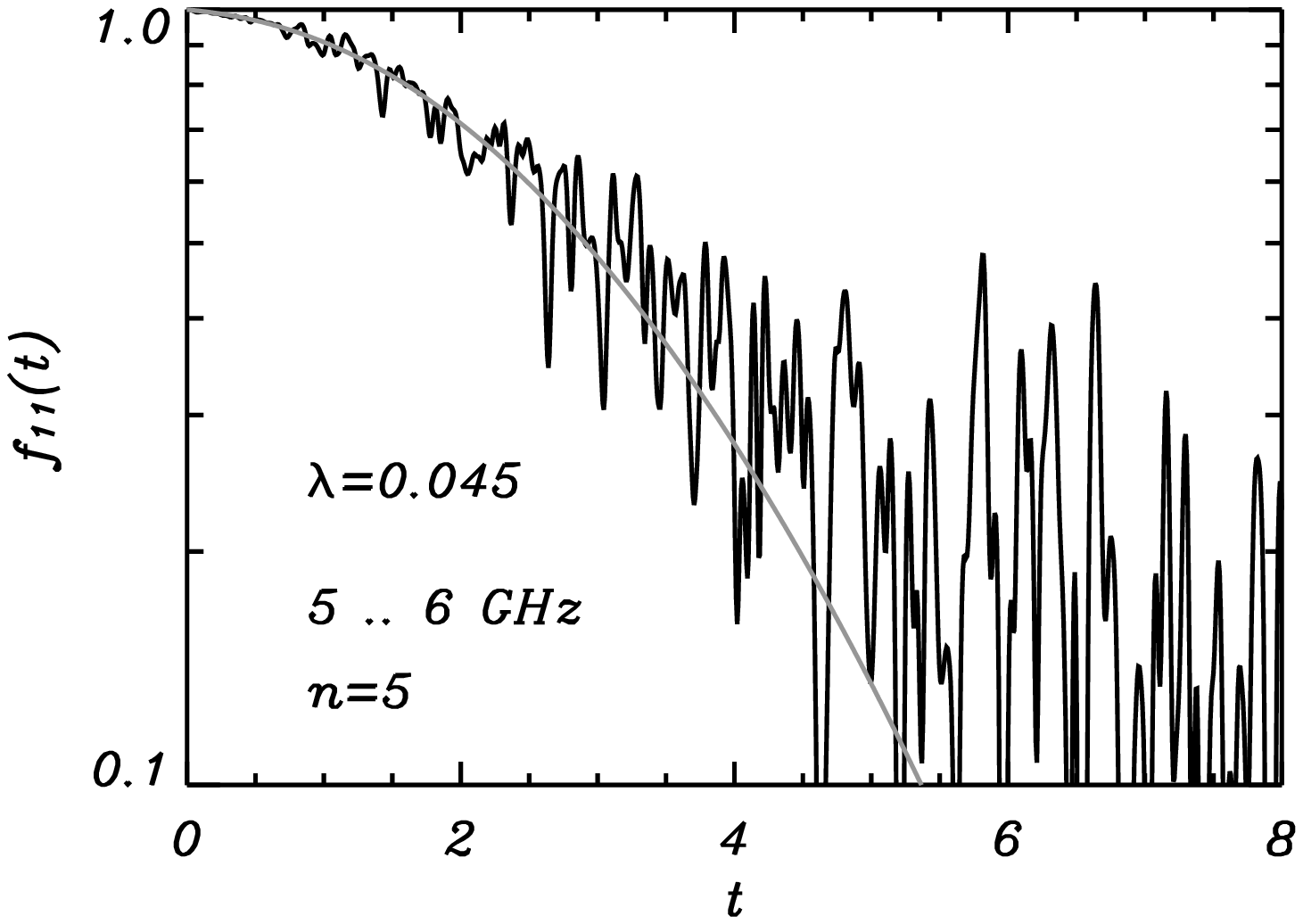}
 \includegraphics[width=0.48\textwidth]{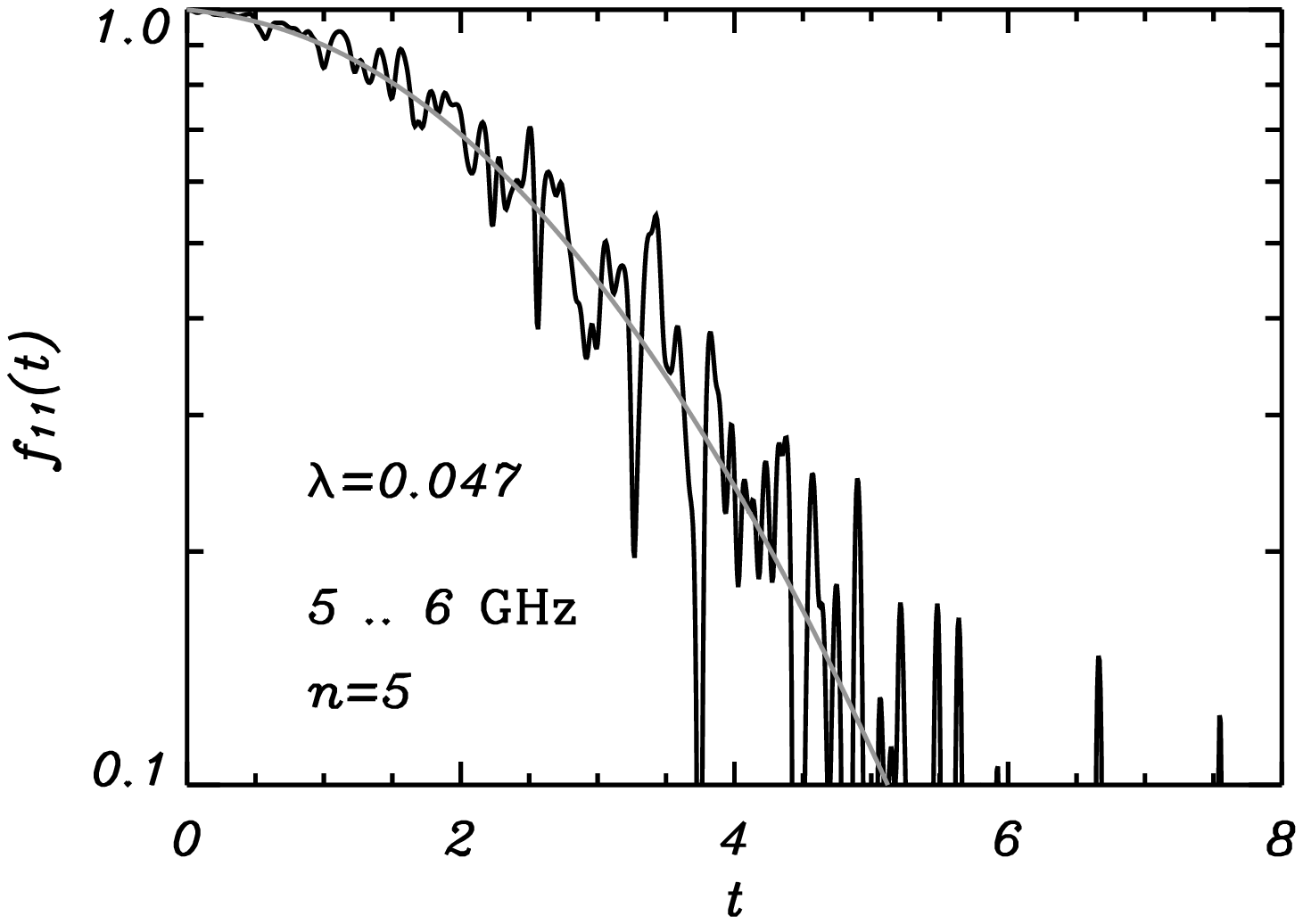} }
 \caption{(top): Logarithmic plot of the correlation function
$\hat C[S^*_{11},S'_{11}]$  for the billiard with bouncing balls
(left), and for the one without bouncing-balls (right). The
experimental results for the autocorrelation are shown in black,
while the correlation of perturbed and unperturbed system are
shown in grey. The smooth solid curve corresponds to the
theoretical autocorrelation function, and the dashed curve to the
product of autocorrelation function and fidelity amplitude. \\
(bottom): Logarithmic plot of the corresponding fidelity
amplitudes. The smooth curve shows the linear-response result.
For the billiard without bouncing balls the perturbation parameter $\lambda$ was obtained from the variance of the level velocities; in the other case it was fitted to the experimental curve. }
\label{fig:auto}
\end{figure}

We start with a discussion of the Fourier transform of the
correlation function $\hat{C}[S^*_{11},S'_{11}](t)$ as given in
equation~(\ref{M:FCfun}). Figure \ref{fig:auto} shows a logarithmic
plot of $\hat C[S^*_{11},S'_{11}](t)$  in comparison to the
autocorrelation function $\hat C[S^*_{11},S_{11}](t)$. The change of
area and surface due to the perturbation, i.\,e. the shift of the
billiard wall, was taken into account by unfolding the spectra to a
mean level distance of one. The frequency window of the Fourier
transforms was 1\,GHz wide, and a Welch filter was applied. The
correlation functions were averaged over an ensemble of 15 billiard
geometries.

The autocorrelation follows the corresponding theoretical curve
nicely, where the parameters for the wall absorption and the
coupling of the antennas had been determined as described in
reference \cite{Sch03a}. Minor deviations from the theoretical
curve can be attributed to the small number of levels in this
frequency interval. Furthermore, small slits between the inserts
of the billiards may act as additional decay channels. This is in
contrast to the more or less homogenous absorption by the cavity
wall, which can be described by infinitely many weak channels
\cite{Sch03a}.

With increasing time, the correlation function $\hat
C[S^*_{11},S'_{11}](t)$ deviates more and more from the
autocorrelation function. This behaviour can be described by a
product of the autocorrelation function and the fidelity amplitude
$f(t)$ of the closed system, thus confirming expression
(\ref{eq:avg_va_wcl}).

As an expression for the fidelity amplitude, we use the one
derived by Gorin et al. \cite{gor04} based on the linear-response
approximation,
\begin{equation}\label{eq:fd_lin}
  f(t)= e^{- 4 \pi^2 \lambda^2 \, C(t)}\,.
\end{equation}
For the Gaussian orthogonal ensemble $C(t)$ is given by
\begin{equation}\label{00aa}
    C(t)= t^2+\frac{t}{2}-\int_0^t \int_0^\tau
b_2(\tau^\prime) \rmd \tau^\prime \rmd \tau \,,
\end{equation}
where $b_2(t)$ is the two-point form factor. Equation
(\ref{eq:fd_lin}) describes the fidelity amplitude both in the
pertubative regime, where the Gaussian decay is dominant, and in
the Fermi golden rule regime, which shows an initial exponential
decay.

In the following we shall study the fidelity amplitude directly as
the deviation from the autocorrelation function as defined in
equation (\ref{eq:def_fscat}) to achieve a more direct comparison
with theory.
%
%
There are two reasons for dividing by the experimental
autocorrelation function: there is no need to fit the
autocorrelation function, and the influence of non-generic
features in the Fourier transform is reduced.

The plots in the lower row of figure \ref{fig:auto} show that this
procedure works very well. For the billiard without bouncing balls 
and the frequency range shown in this
figure, the perturbation strength was determined directly from the
measured spectra via the variance of the level velocities.
Thus we can describe the experimental results for the fidelity
amplitude by the linear-response expression without any free
parameter.

As expected, the billiard with bouncing balls shows systematic
deviations from the random matrix prediction. Only for
small times $t$ we find good agreement. We therefore
concentrate on the results of the billiard without bouncing balls in
the following subsections.

\subsection{Agreement with the linear-response prediction}

\begin{figure}
\centerline{
\includegraphics[width=0.48\textwidth]{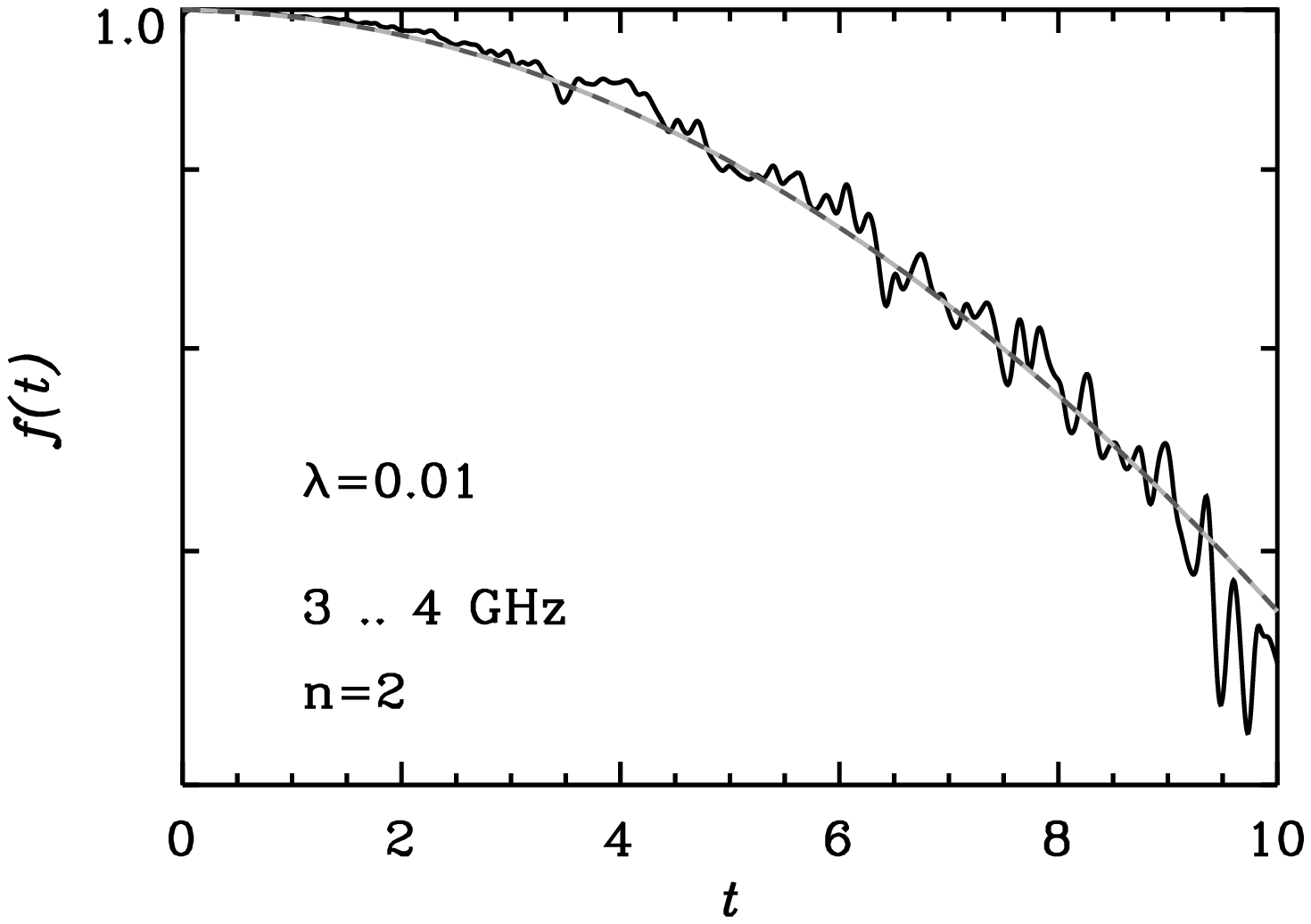}
 \includegraphics[width=0.48\textwidth]{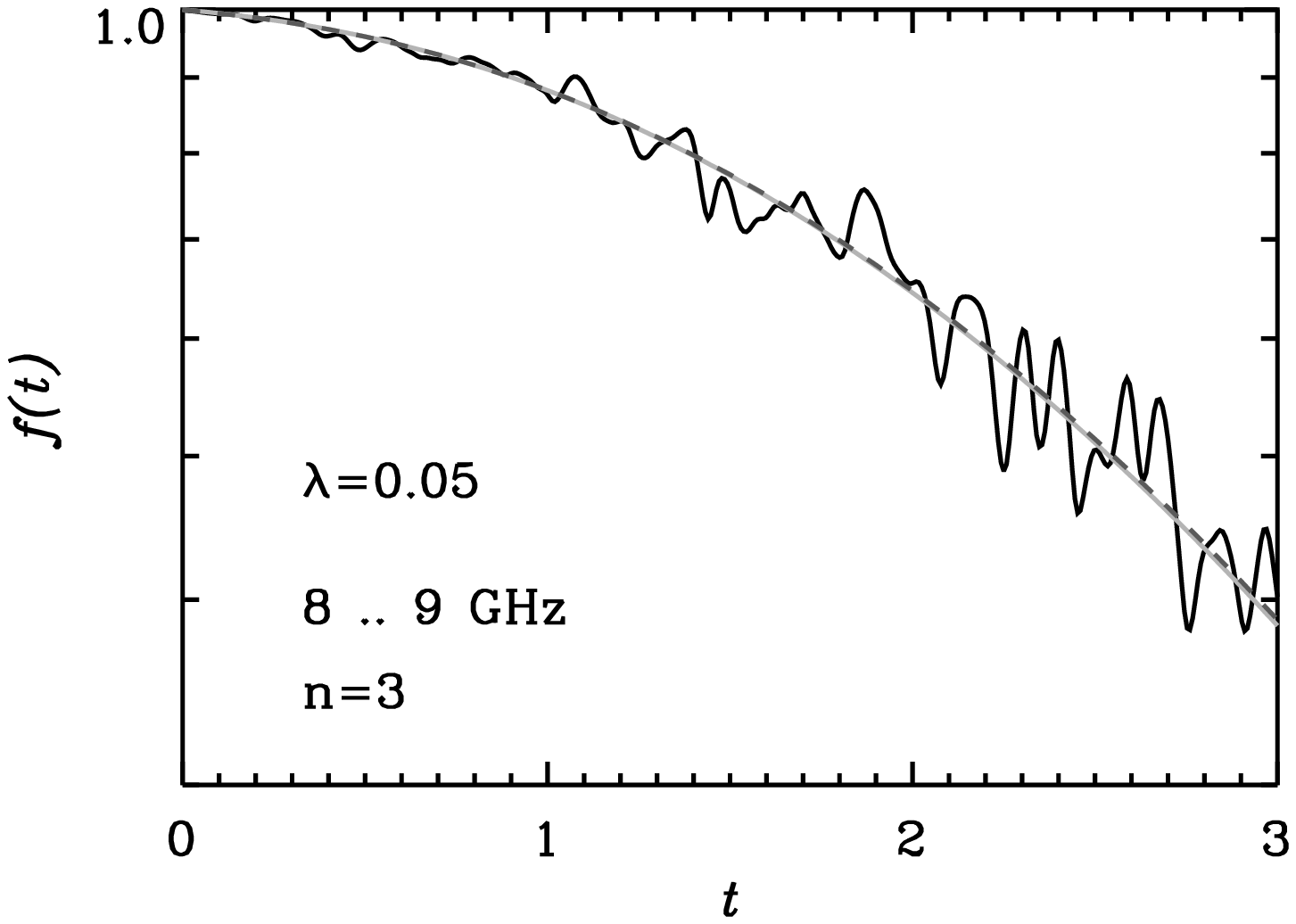} }

\centerline{
\includegraphics[width=0.48\textwidth]{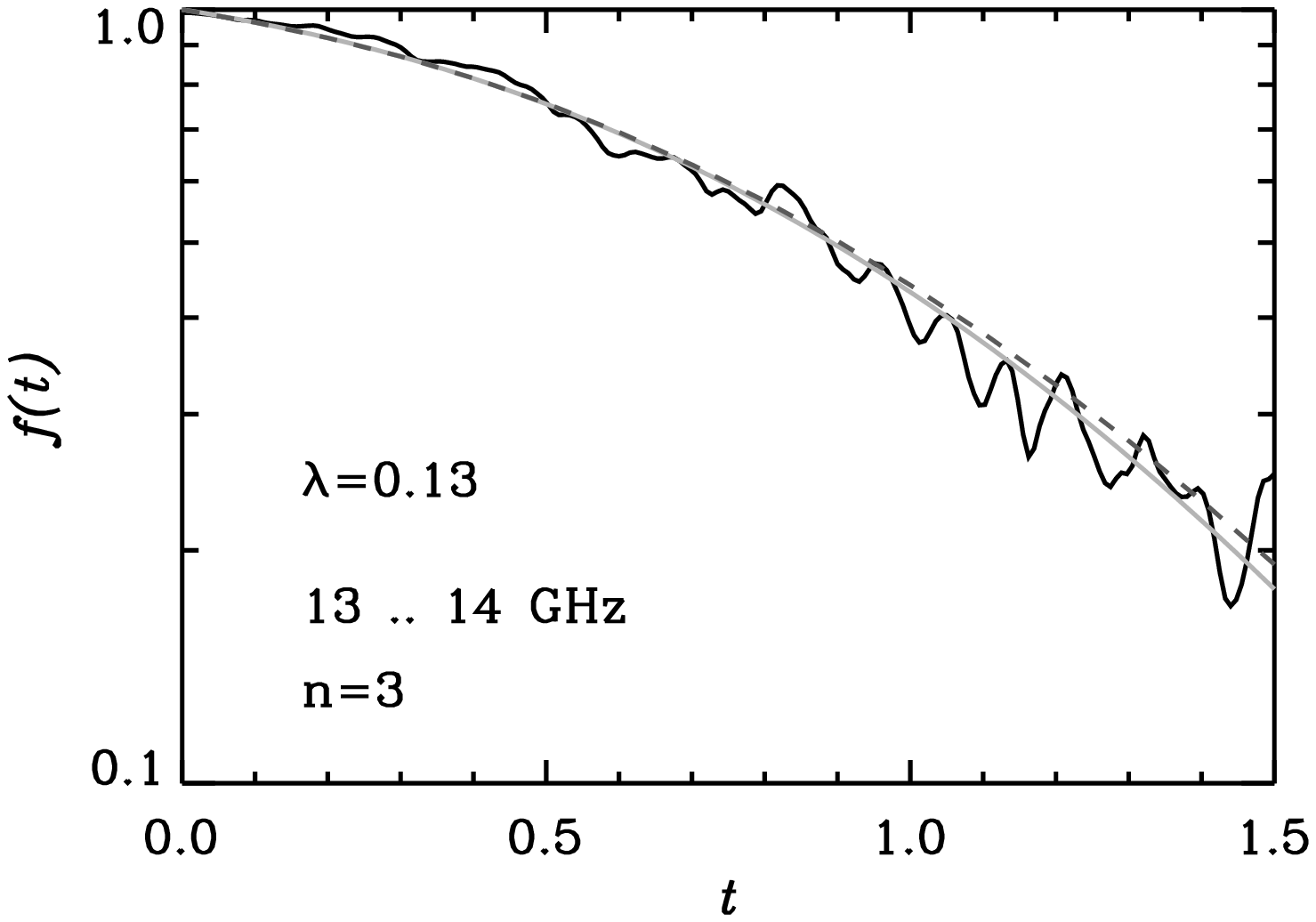}
 \includegraphics[width=0.48\textwidth]{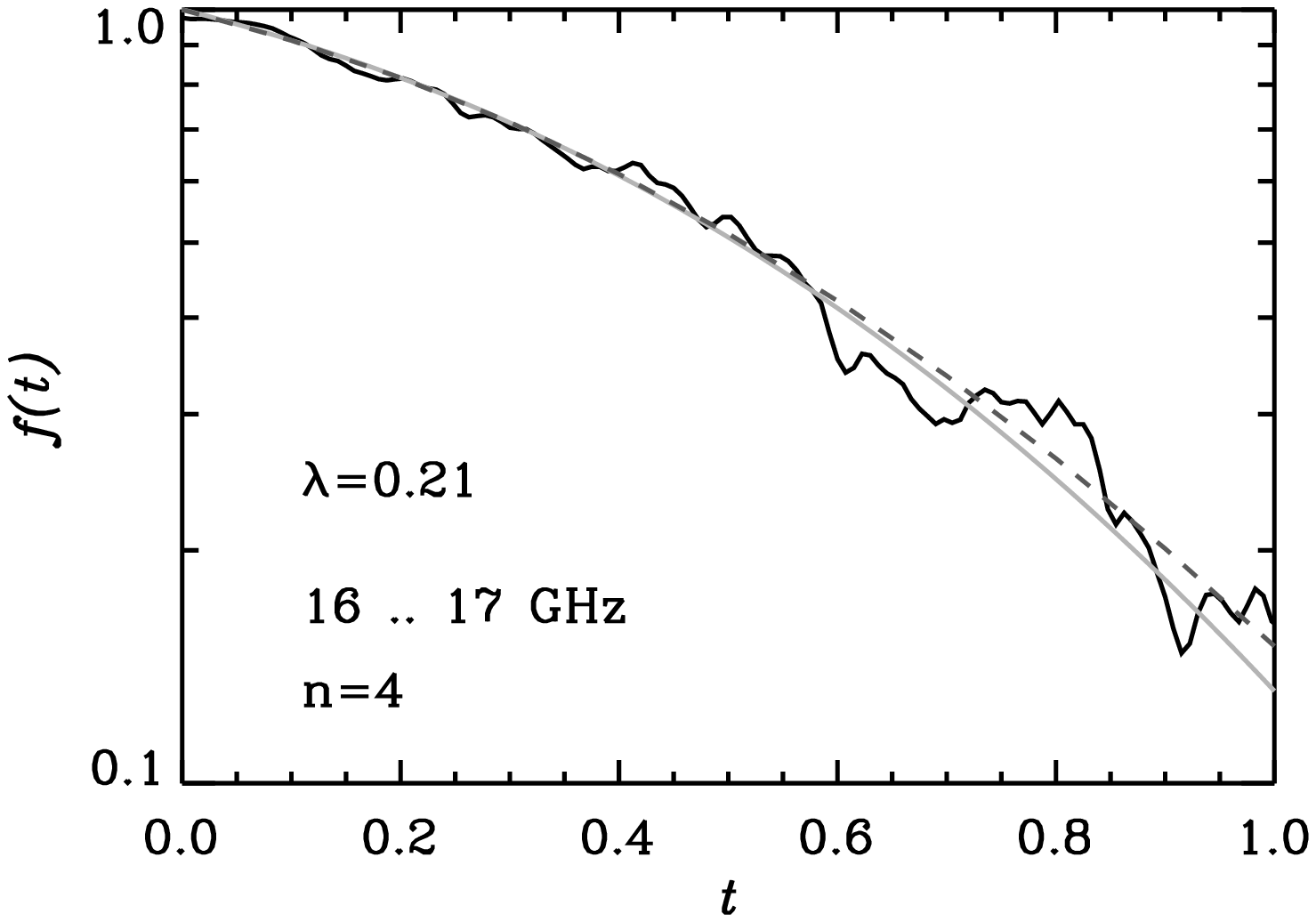} }
\caption{Fidelity amplitude of the billiard without bouncing
balls, obtained by superimposing results for $f_{11}$, $f_{22}$
and $f_{12}$. The smooth solid line shows the linear-response
result (\ref{eq:fd_lin}), while the dashed line shows the exact
result \cite{stoe}. The perturbation parameter $\lambda_{\rm exp}$
has been fitted to each experimental curve.} \label{fig:fidelity}
\end{figure}

In our experiment the values for the perturbation parameter
$\lambda$ vary from $\lambda=0.01$ for $n=1$ and $\nu=3..4$\,GHz
up to $\lambda=0.5$ for $n=10$ and $\nu=17\ldots 18$\,GHz.
%
Figure \ref{fig:fidelity} shows the fidelity amplitude for four
different frequency windows. The perturbation parameter $\lambda$
has been fitted to the experimental curves. To improve statistics,
experimental results for $f_{11}$, $f_{22}$ and $f_{12}$ have been
superimposed. The individual curves for these quantities were not
discernible within the limit of error.

For small perturbation strengths, the linear term in the
exponential is still close to one and thus we observe essentially
a Gaussian decay of the fidelity amplitude, as seen in figure
\ref{fig:fidelity} for $\lambda_{\rm exp}=0.01$. With increasing
perturbation strength the linear term is getting more pronounced,
leading to an exponential decay for small times. However, for
larger times the Gaussian decay again becomes dominant.

In the range accessible to our experiment (limited by the small
ensemble) we find good agreement with the linear-response
prediction of the random matrix model.
For the strongest perturbation strength realized in our
experiment, deviations of the linear-response result from the
exact result for the Gaussian ensemble~\cite{stoe} become
noticeable only at times where $f(t)$ has decayed already 
below $10^{-1}$. These deviations could not be
detected by our experiment.


In~\cite{Weaver} Lobkis and Weaver report on a very similar
experiment in elastodynamic billiards. There, the principal quantity
of interest is called distortion, denoted by $D(t)$. In our
terminology, it is the logarithm of the scattering fidelity: $D(t)=
-\ln f_{aa}(t)$. Interestingly, the experiment itself is performed
in the time domain. A short acoustic pulse is coupled into the
sample via a pointlike pin transducer, and with the same device the
time signal corresponding to the response of the system is measured.
Cross-correlating two such signals, measured at slightly different
temperature allows to extract the scattering fidelity, where the
perturbation is due to the temperature change. In~\cite{Wcomm} 
the experimental data from~\cite{Weaver} are reanalyzed, showing
again an excellent agreement with the random matrix prediction,
derived in~\cite{gor04}.

\subsection{Scaling behavior of the perturbation strength}

The experimental fidelity amplitude was studied for 10 different
shifts of the billiard wall ($\Delta l=n \cdot 0.2$\,mm, with
$n=1\ldots10$), and a frequency window of width 1\,GHz was moved
through the spectrum. The perturbation strength $\lambda^2$
entering the fidelity amplitude~(\ref{eq:fd_lin}) was fitted to
the experimental results.

\begin{figure}
\centerline{
\includegraphics[width=0.48\textwidth]{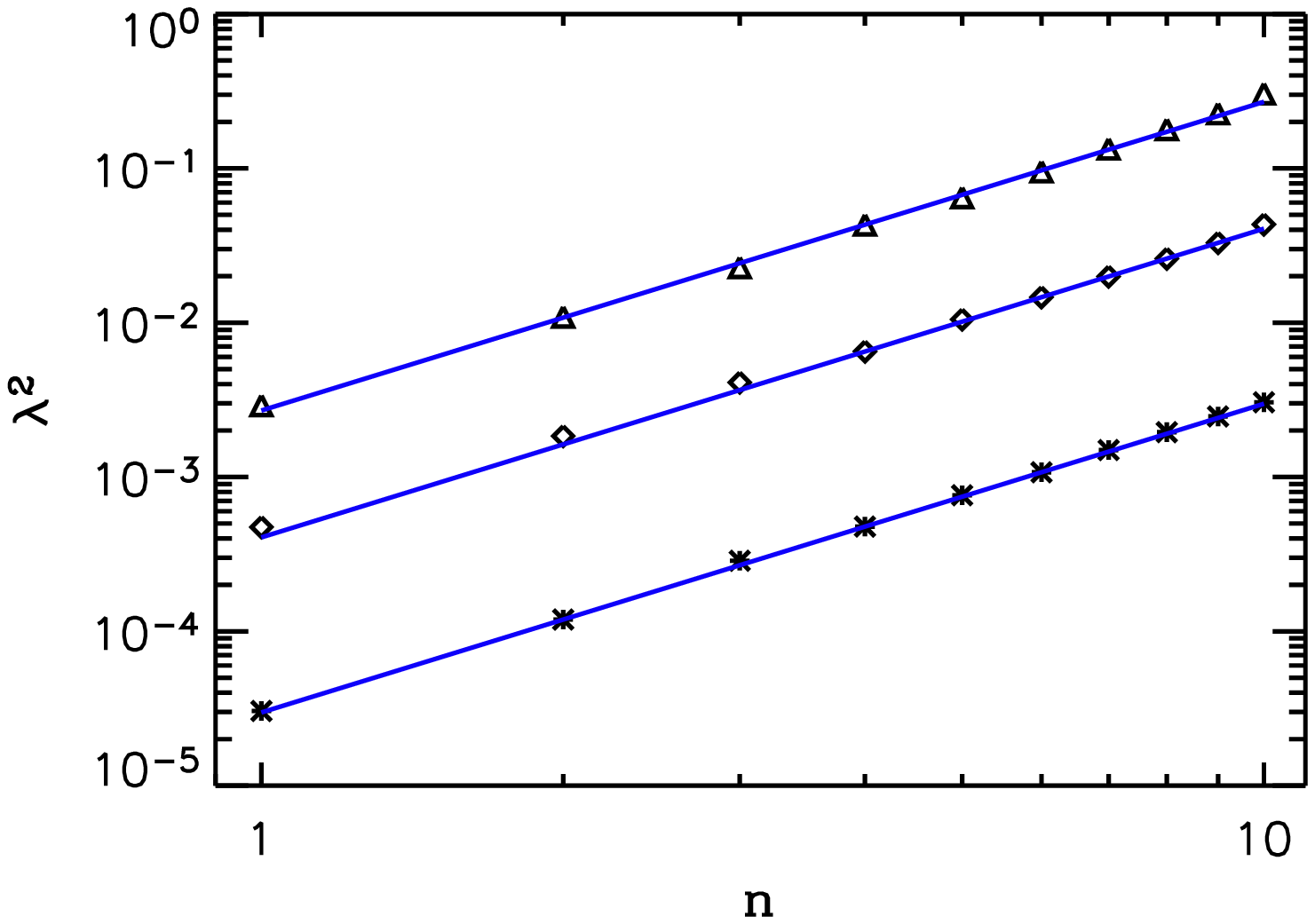}
 \includegraphics[width=0.48\textwidth]{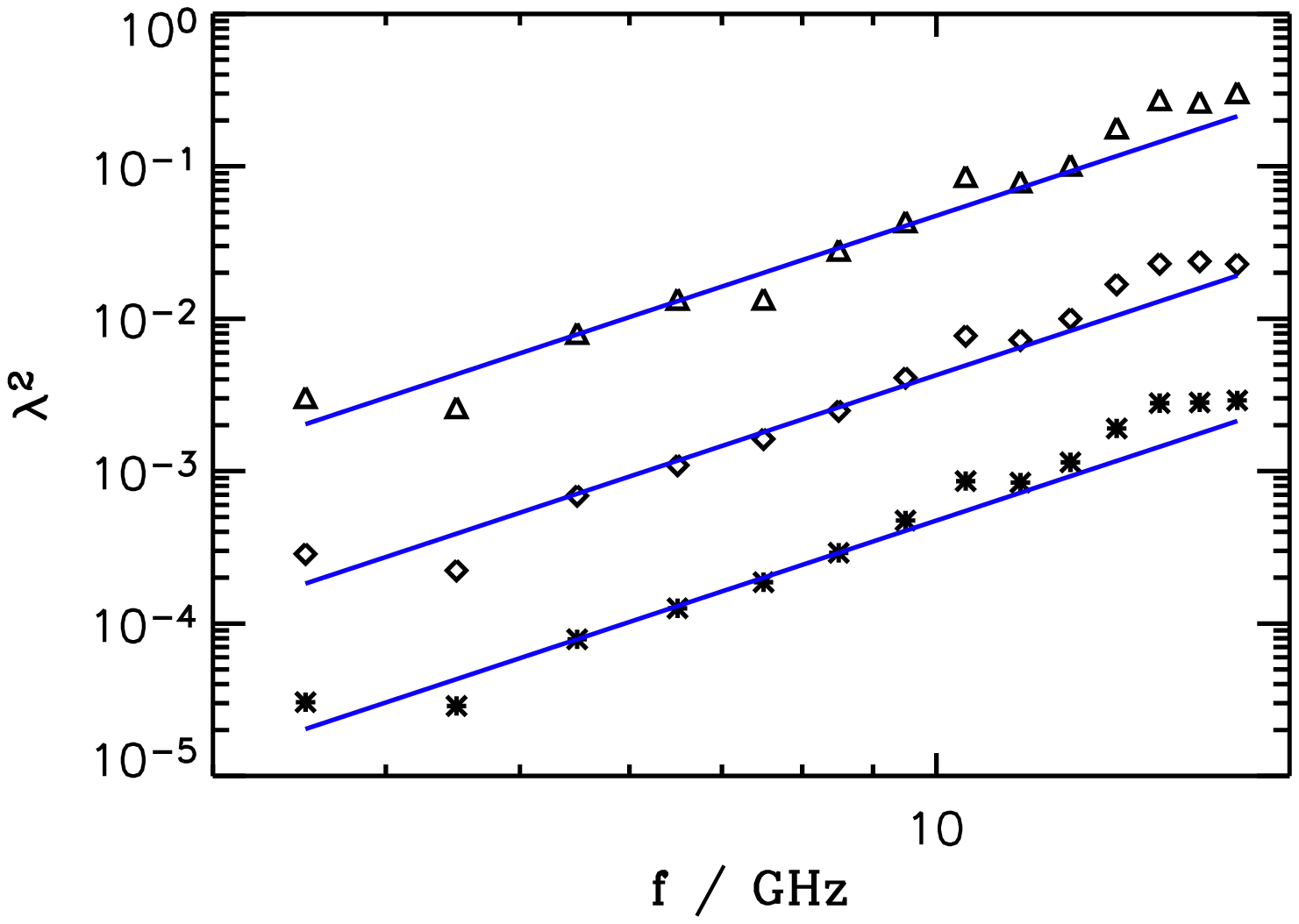} }
\caption{Perturbation strength $\lambda^2$ as a function of $n$
(left), and as a function of frequency $\nu$ (right). The slopes
of the straight lines are $2$ in the left, and $3$ in the right
hand figure. The results are shown for the billiard without
bouncing balls.} \label{fig:scale}
\end{figure}

Figure \ref{fig:scale}(a) shows the experimental perturbation
strength $\lambda_{\rm exp}^2$ in dependence of the number of
steps $n$ for three different frequency regimes. We observe an
excellent agreement with the scaling $\lambda^2 \propto n^2$ as
predicted from equation (\ref{eq:scaling}). The experimental
results for $\lambda_{\rm exp}^2$ in dependence of the frequency
range are shown in figure \ref{fig:scale}(b). They do not look quite as
nice as the previous results, but they still confirm the scaling
$\lambda^2 \propto \nu^3$.

\begin{figure}
\centerline{
\parbox{0.48\textwidth}{
}
\parbox{0.48\textwidth}{
\includegraphics[width=0.48\textwidth]{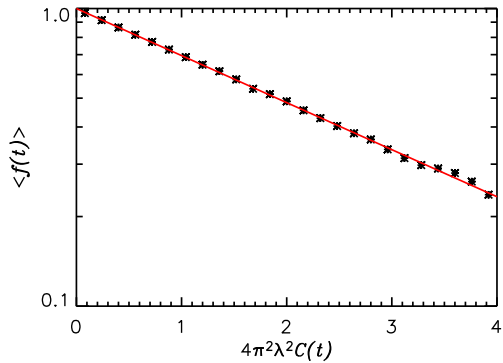} }
} \caption{Average of the experimental fidelity amplitude on a
rescaled axis $x=4\pi^2\lambda^2C(t)$. The solid line corresponds to
$g(x)=\exp{(-\alpha x)}$ with $\alpha=\lambda_{\rm
exp}^2/\lambda^2=0.36$. } 
\label{fig:scalingall}
\end{figure}

In view of the good agreement with the predicted scaling of the
perturbation strength, we can now test the time dependence implied
by $C(t)$ with better statistics by averaging over all data after
rescaling: In figure~\ref{fig:scalingall}, these averages are
plotted against $4\pi^2\lambda^2\, C(t)$ and we see that the
linear response solution of the random matrix model~\cite{gor04}
describes all data very well. However, the slope of the resulting
curve is not in accordance with the prediction and yields a
$\lambda_{\rm exp}^2$ deviating from the theoretical expectation
of equation (\ref{eq:scaling}) by a factor of $0.36$.

This deviation is caused by the fact that we are far from the
semi-classical limit, for which the expression for $\lambda^2$ was
derived. This is in accordance with numerical calculations for the
Sinai billiard done by H. Schanz in G\"ottingen \cite{schanz}.
In cases where we determined the  variance of level velocities
directly from the measured spectra, we found the same deviation
from equation (\ref{eq:scaling}). This shows that the experiment
can be described in a self-consistent way.

\subsection{Influence of bouncing-ball modes}

To study the frequency dependence of the perturbation strength in
some more detail, we took a smaller frequency window of 0.5\,GHz
for the Fourier transform and moved it in finer steps through the
whole frequency range.

\begin{figure}
\centerline{ \includegraphics[width=0.48\textwidth]{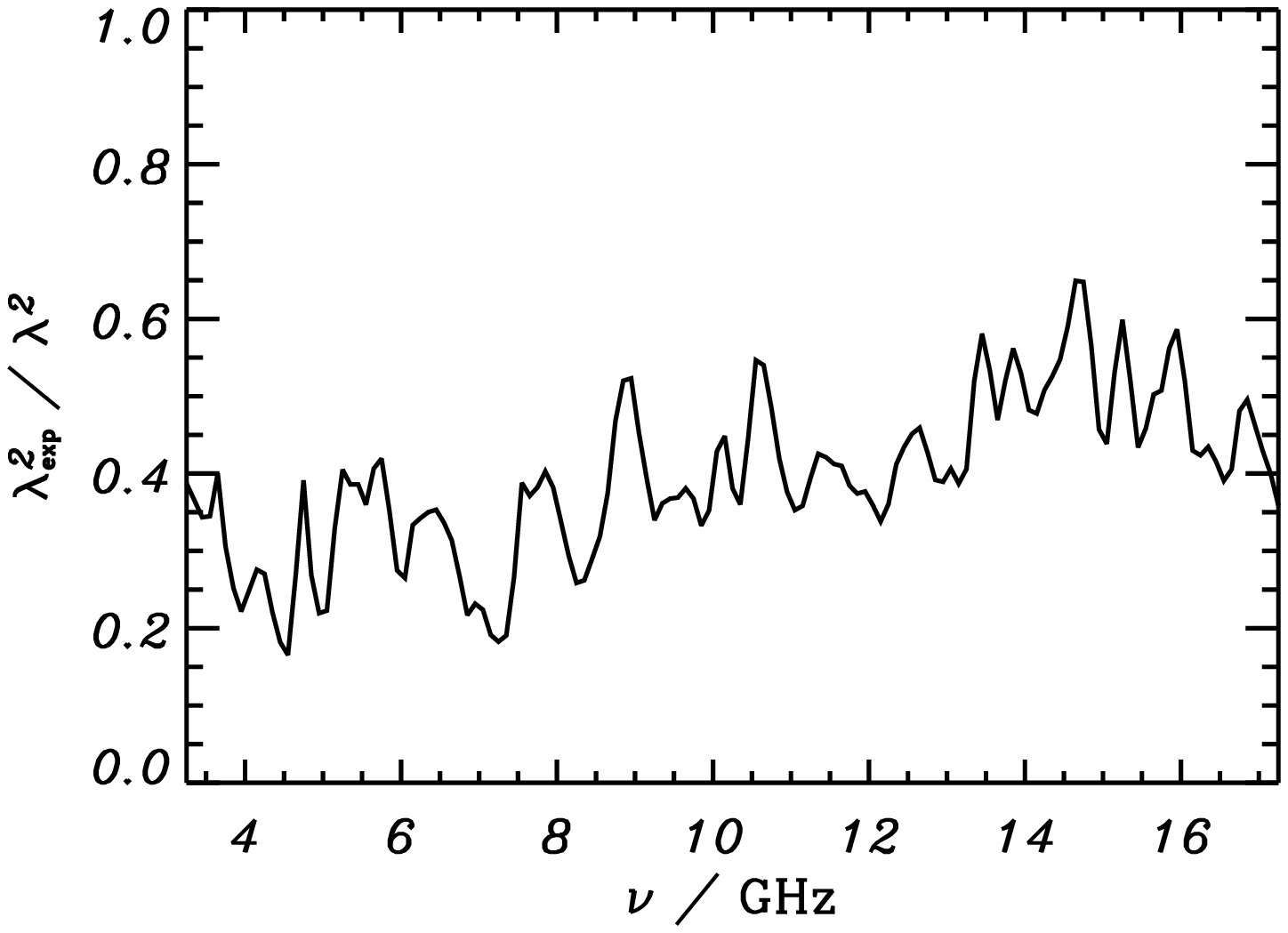}
 \includegraphics[width=0.48\textwidth]{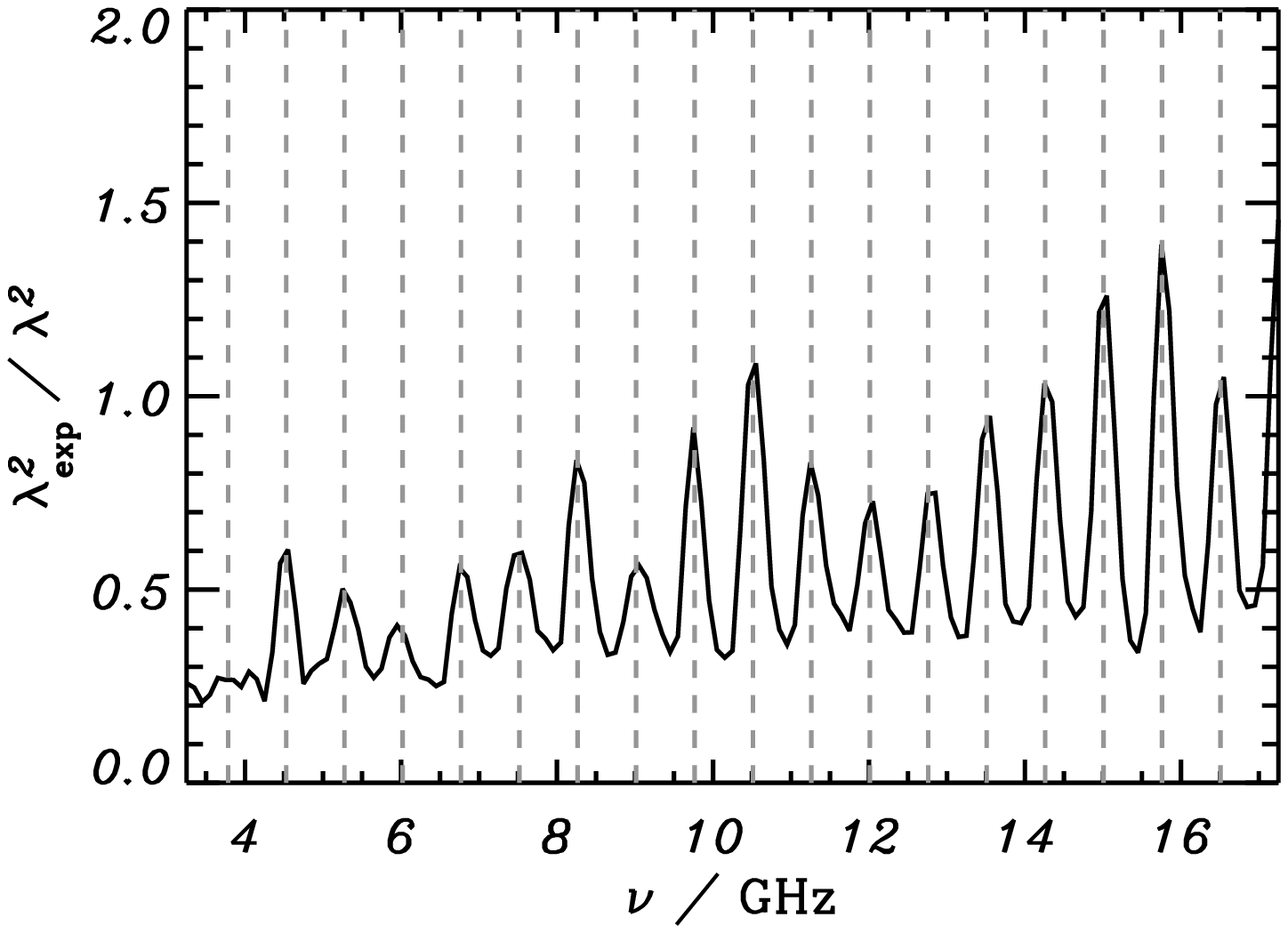} }
\caption{$\lambda_{\rm exp}^2/\lambda^2$ as a function of frequency
$\nu$. Results for the billiard without (left) and with bouncing
balls (right). The dashed lines correspond to the
eigenfrequencies of the associated bouncing-ball states. } 
\label{fig:correction}
\end{figure}

In Figure \ref{fig:correction} the ratio $\lambda_{\rm
exp}^2/\lambda^2$ is plotted both for the billiard with and
without bouncing balls, where $\lambda^2$ is the theoretical value
of the perturbation strength according to equation
(\ref{eq:scaling}).

In the right hand plot, the frequencies of the vertical bouncing
ball resonances are plotted as vertical lines, revealing the origin
of the peaks in the frequency dependence of $\lambda_{\rm exp}^2$.
The vertical bouncing ball states correspond to standing waves between the 
long sides of the billiard and remain nearly unaffected by the
perturbation. However, the unfolding of each spectrum takes the change of
area into account, and thus introduces a constant drift to these
resonances. This leads to a faster decay of the fidelity, thus
resulting in the strong peaks visible in the figure.

Note  that the result for the billiard without bouncing balls
still shows fluctuations. The influence of the eigenmodes in the
individual frequency window is thus visible for this billiard as
well. However, there are no systematic peaks like the ones for the
bouncing-ball modes.

\section{Conclusions}

It was shown in this paper that the scattering fidelity
$f_{ab}(t)$ is an easily accessible quantity which in the
weak-coupling limit approaches the ordinary fidelity amplitude
$f(t)$. It is stressed that it is not necessary to vary the
antenna position for this purpose.
For integrable systems, however, an average over the antenna
positions is indispensable to obtain $f(t)$. Therefore a
determination of the fidelity $f(t)$ from the scattering fidelity
$f_{ab}(t)$ as illustrated here, is probably not feasable in that
case.

All results of the present paper could be described within the
limits of the exponentiated linear-response approximation \cite{gor04}, 
allowing a nice scaling of all data onto one single curve. Deviations of
the decay constant of the such scaled fidelity from the predicted
value could be quantitatively traced back to limits of
the semiclassical approximation.

A general perturbation of the scattering matrix, is not necessarily
restricted to the internal dynamics, as it may also change the
coupling to the channels. For practical reasons, the present
experimental setup was not well suited to study such a situation,
and we are planning a further experiment to study scattering
fidelity as a function of coupling strength.

\ack

T Prosen, U Kuhl and H Schanz are thanked for helpful discussions
taking place in part on occasion of a workshop at the Centro
Internacional de Ciencias in Cuernavaca, Mexico. We thank R L Weaver for
drawing our attention to reference~\cite{Weaver}. 
The experiments were supported by the Deutsche Forschungsgemeinschaft, and 
T. H. S. acknowledges support under grants CONACyT  \# 41000-F and UNAM-DGAPA 
IN-101603.

\begin{appendix}

\section{\label{app} Scattering fidelity in the rescaled Breit-Wigner 
   approximation}

We assume that the
spectrum of the unperturbed closed system (Hamiltonian $H_\inter$) may be taken
from the Gaussian orthogonal ensemble (GOE), and consider averages over
that ensemble. In the rescaled Breit-Wigner approximation~\cite{Gor02a}, the
correlation function (\ref{eq:tauprime}) reads
\begin{eqnarray}
\fl \lla \hat S_{ab}(t)^*\, \hat S'_{ab}(t)\rra & = & 4\pi^2\theta(t)
   \lla \sum_{jk} \la a|V|j\ra \, \la j|\,
   \rme^{2\pi\rmi (E_j +\rmi\Gamma_j/2) t}\, |j\ra \; \la j|V|b\ra \right.
\nonumber\\
&&\qquad\qquad\times\; \left. \la b|V|k'\ra\; \la k'|\,
   \rme^{2\pi\rmi (E'_j -\rmi\Gamma'_j/2) t}\, |k'\ra\; \la k'|V|a\ra
   \rule{0pt}{15pt}\rra \; ,
\end{eqnarray}
where we have used Dirac's notation for the eigenbases $\sum_j |j\ra
\, \la j|$ and $\sum_k |k'\ra \, \la k'|$ of $H_\inter$ and
$H'_\inter$, respectively. The numbers $\Gamma_j$ and $\Gamma'_j$
denote the diagonal element of the anti-Hermitean part of $H_{\rm
eff}$ and $H'_{\rm eff}$ in the respective basis. We will relate
this correlation function to the fidelity amplitude of the closed
system $H_\inter$, or $H'_\inter$ (when perturbed). To this end, we
need to restrict ourselves to the perturbative regime. In this case,
$|j'\ra = |j\ra$
and $\Gamma_j' = \Gamma_j$, so that
\begin{eqnarray}
\fl && \lla \hat S_{ab}(t)^*\, \hat S'_{ab}(t)\rra = 4\pi^2\theta(t)
   \sum_{jk} \lla V_{ja}^*\, V_{jb}\, V_{kb}^*\, V_{ka}\;
   \rme^{-\pi (\Gamma_j+\Gamma_k) t}\; \rme^{-2\pi\rmi (E'_k - E_j) t} \rra
\nonumber\\
\fl &&\qquad = \sum_k \lla \rme^{-2\pi\rmi (E'_k - E_k) t} \rra \sum_j
   \lla V_{ja}^*\, V_{jb}\, V_{kb}^*\, V_{ka}\;
   \rme^{-\pi (\Gamma_j+\Gamma_k) t}\; \rme^{-2\pi\rmi (E_k - E_j) t}
   \rule{0pt}{13pt}\rra \; .
\end{eqnarray}
The separation of averages in the second line is possible in the
perturbative regime, only. There $E'_k-E_k$ is just the diagonal
element of the perturbation matrix and hence statistically
independent of the unperturbed system. 
The average $\la\exp[-2\pi\rmi (E'_k-E_k) t]\ra$ gives
the fidelity amplitude, independent of the index $k$, while the rest
is just the autocorrelation function of the S-matrix element
$S_{ab}$.
\begin{eqnarray}
\label{eq:corrfd} \lla \hat S_{ab}(t)^*\, \hat S'_{ab}(t)\rra =
f(t)\;  \lla |\hat S_{ab}(t)|^2\rra \; .
\end{eqnarray}
\end{appendix}


\section*{References}

\begin{thebibliography}{10}

\bibitem{Peres84}
Peres A
\newblock 1984 {\em Phys. Rev.} A {\bf 30}~1610

\bibitem{nielsen}
Nielsen M A and Chuang I L
\newblock 2000 {\em Quantum Computation and Quantum Information}
\newblock (Cambridge, Cambridge University Press)

\bibitem{pro03b}
Prosen T, Seligman T H  and \v{Z}nidari\v{c} M
\newblock 2003 {\em Prog. Theor. Phys. Suppl.} {\bf 150}~200

\bibitem{GPSS04}
Gorin T, Prosen T, Seligman T H and Strunz W T
\newblock 2004 {\em Phys. Rev.} A {\bf 70}~042105

\bibitem{GCZ97}
Gardiner S A, Cirac J I and Zoller P
\newblock 1997 {\em Phys. Rev. Lett.} {\bf 79}~4790

\bibitem{bienert}
Haug F, Bienert M, Schleich W P, Seligman T H and Raizen M G
\newblock 2005 {\em Phys. Rev. A} {\bf 71}~043803

\bibitem{SSGS04}
Sch\" afer R, St\" ockmann H-J, Gorin T and Seligman T H
\newblock 2004 {\em nlin.CD}/0412053

\bibitem{Weaver}
Lobkis O I and Weaver R L
\newblock 2003 {\em Phys. Rev. Lett.} {\bf 90}~254302

\bibitem{nemes}
Nemes M C and Seligman T H
\newblock 1980 {\em Z. Phys. A} {\bf 295}~243

\bibitem{jung}
Jung C and Seligman T H
\newblock 1997 {\em Phys. Rep.} {\bf 285}~77

\bibitem{gor04}
Gorin T, Prosen T  and Seligman T H
\newblock 2004 {\em New J. Phys.} {\bf 6}~20

\bibitem{FTbook}
Brigham E O
\newblock 1974 {\em The Fast Fourier Transform}
\newblock (New York: Prentice Hall)

\bibitem{EngWei73}
Engelbrecht, C A and Weidenm\" uller H A
\newblock 1973 {\em Phys. Rev. C} {\bf 8}~859

\bibitem{Stoe99}
St\"ockmann H J
\newblock 1999 {\em Quantum Chaos - An Introduction}
\newblock  (Cambridge: University Press)

\bibitem{stoe}
St\"ockmann H J and Sch\"afer R
\newblock 2004 {\em New J. Phys.} {\bf 6}~199

St\"ockmann H J and Sch\"afer R
\newblock 2004 {\em nlin.CD}/0409021

\bibitem{Wcomm}
Gorin T, Seligman T H and Weaver R L
\newblock 2005 unpublished

\bibitem{Kuh00b}
Kuhl U, Persson E, Barth M  and St\"ockmann H J
\newblock 2000 {\em Eur. Phys.~J. B} {\bf 17}~253

\bibitem{Mah69}
Mahaux C and Weidenm\"uller H A
\newblock 1969 {\em Shell-Model Approach to Nuclear Reactions}
\newblock  (Amsterdam: North-Holland)

\bibitem{Sch03a}
Sch\"afer R, Gorin T, Seligman T H  and St\"ockmann H J
\newblock 2003 {\em J. Phys. A} {\bf 36}~3289

\bibitem{Gor02a}
Gorin T and Seligman T H
\newblock 2002 {\em Phys. Rev. E} {\bf 65}~026214

\bibitem{Berry77}
Berry M V 1977 {\em J. Phys. A} {\bf 10}~2083

Berry M V 2002 {\em J. Phys. A} {\bf 35}~3025

\bibitem{Leb99}
Leb{\oe}uf P and Sieber M
\newblock 1999 {\em Phys. Rev. E} {\bf 60}~3969


\bibitem{schanz}
Schanz H
\newblock 2004 {\em Private communication}

\end{thebibliography}

\end{document}